\documentclass[conference]{IEEEtran}
\IEEEoverridecommandlockouts
\usepackage{cite}
\usepackage{amsmath,amssymb,amsfonts}
\usepackage{algorithmic}
\usepackage{graphicx}
\usepackage{textcomp}
\usepackage{xcolor}
\def\BibTeX{{\rm B\kern-.05em{\sc i\kern-.025em b}\kern-.08em
    T\kern-.1667em\lower.7ex\hbox{E}\kern-.125emX}}
\DeclareMathOperator{\Tr}{Tr}

\begin{document}

\title{URLLC with Coded Massive MIMO via\\ Random Linear Codes and GRAND}


\author{
    \IEEEauthorblockN{Sahar Allahkaram\IEEEauthorrefmark{1}, 
     Francisco A. Monteiro\IEEEauthorrefmark{1}, 
	 and Ioannis Chatzigeorgiou\IEEEauthorrefmark{2}}
    \IEEEauthorblockA{\IEEEauthorrefmark{1}Instituto de Telecomunicações, and ISCTE - Instituto Universitário de Lisboa, Lisbon, Portugal}
    \IEEEauthorblockA{\IEEEauthorrefmark{2}School of Computing and Communications, Lancaster University, Lancaster, United Kingdom }
   
emails: \{sahar.allahkaram, francisco.monteiro\}@lx.it.pt, i.chatzigeorgiou@lancaster.ac.uk

}
\maketitle

\begin{abstract}
A present challenge in wireless communications is the assurance of ultra-reliable and low-latency communication (URLLC). While the reliability aspect is well known to be improved by channel coding with long codewords, this usually implies using interleavers, which introduce undesirable delay. Using short codewords is a needed change to minimizing the decoding delay. This work proposes the combination of a coding and decoding scheme to be used along with spatial signal processing as a means to provide URLLC over a fading channel.
The paper advocates the use of random linear codes (RLCs) over a massive MIMO (mMIMO) channel with standard zero-forcing detection and guessing random additive noise decoding (GRAND). The performance of several schemes is assessed over a mMIMO flat fading channel. The proposed scheme greatly outperforms the equivalent scheme using 5G's polar encoding and decoding for signal-to-noise ratios (SNR) of interest. While the complexity of the polar code is constant at all SNRs, using RLCs with GRAND achieves much faster decoding times for most of the SNR range, further reducing latency.

\end{abstract}


\section{Introduction}
The physical layer has much to contribute to the goal of wireless ultra-reliable and low-latency communication (URLLC), whose major goals are to reduce latency to $1\mathrm{~ms}$ and simultaneously ensure at least 99.999 \% reliability \cite{Nouri2020}.

Noise-guessing decoding was very recently proposed as a universal decoder for codes having codewords of moderate length or sufficiently high rate, which are ideal for applications in wireless URLLC, where the codewords are desirably short. The technique, dubbed guessing random additive noise decoding (GRAND), allows ditching interleavers and eliminating the decoding delay bottleneck they imposed \cite{An2022}. Given that the entropy of the noise is much smaller than the entropy of the codewords, decoding of the noise pattern that affected a codeword greatly diminishes the complexity of maximum likelihood (ML) decoding. Moreover, because it is a universal decoder, it opened doors to using random linear codes (RLCs), which are able to attain the maximum rate of the finite-blocklength regime, precisely the one of interest in URLLC.
The combination of RLC encoding with GRAND is a perfect fit for URLLC, where the flexibility of choice of the codewords' lengths and rates of a RLC greatly exceeds any limitation regarding their length.

As it is well known, massive multiple-input multiple-output (mMIMO) allowed the very high spectral efficiencies in 5G due to spatial multiplexing.
In order to cater for both objectives in URLLC, this paper proposes coded mMIMO links using a RLC and GRAND. GRAND has been recently proposed for single-input single-output (SISO) flat Rayleigh channels and shown to outperform Bose–Chaudhuri–Hocquenghem (BCH) codes using the Berlekamp-Massey (B-M) decoder \cite{Abbas}. To the best of our knowledge, the use of RLCs and GRAND is for the first time proposed and assessed in the context of a MIMO fading channel.
 
\section{Random Linear Codes}
Shannon's random code-construction \cite{Shannon1948} provably reaches capacity when the length of the codewords tends to infinity and the decoder performs ML decoding.
However, selecting the codebook members randomly leads to two different problems: i) a storage problem, given that all codewords would have to be stored both at the encoding side and at the decoding side, and ii) a decoding complexity problem given that, when applying the ML principle, a corrupted codeword needs to be compared with all codewords in the codebook in order to find the most likely one.

The storage problem posed by Shannon's construction has been overcome by RLCs, because, as in the case of any linear block code, the generator matrix $\mathbf{G}$ encapsulates a very short description of a code, and is the only piece of information needed to be stored.

RLCs are known to be capacity-achieving in the binary symmetric channel (BSC), \cite{Shannon1948, Gallager1973}, in the asymptotic regime. Notably, they also attain the maximum rate possible in the \textit{finite-blocklength regime} \cite{Polyanskiy_2010,Polyanskiy_2014}, the one of interest for practical URLLC. Furthermore, RLCs can be constructed with \textit{any size and any rate}, and thus having those degrees of freedom is a major practical advantage for most engineering applications.

\section{Guessing Random Additive Noise Decoding}
Albeit classical RLCs are known to be capacity-achieving, they do not offer a solution to the remaining problem of having a practical decoder.
A practical solution for the RLC's decoding conundrum only recently emerged with the advent of GRAND-based algorithms \cite{Duffy2019, Duffy2022, An2022}.
GRAND focuses on guessing the noise that corrupted the transmitted codeword, rather than exhaustively going through all the possible codewords, and is proven to still lead to ML decoding \cite{Duffy2019}. GRAND is a \textit{universal} decoder, enabling the decoding of \textit{any} block code of moderate length and with a sufficiently high code rate, and can be used whether the code is random or has some mathematical structure (e.g. polar codes \cite{Duffy2020,Duffy2021}, BCH codes\cite{Abbas, Duffy2021}, Hamming codes, etc), being a binary or multi-level code. Notably, the solely requirement for GRAND is that a membership test exists to detect if a received word is a valid codeword. In the case of RLCs, the test relies on the codeword's syndrome.

Because GRAND is a universal decoder, it opened doors to using the capacity-achieving RLCs for which no practical decoder was yet available. Recent research has shown that RLCs supersede the performance of polar codes of the same length, even with short codewords \cite{An2022, Duffy2020}.
However, while off-the-shelf polar codes do not exist for any wanted length or desired rate, RLCs can be constructed with any desired \textit{blocklength} and \textit{rate}.

For uniform-at-random RLCs, encoding $k$ bits onto $n$, with rate $R=k/n$, where the codewords are chosen uniformly at random from the set (or space) where $2^n$ possible words exist, out of which only $2^k=2^{nR}$ are valid codewords. The distance between any two codewords is also uniformly distributed. As $n\rightarrow \infty$, GRAND permits a successful decoding in  $2^n/2^{nR}=2^{n(1-R)}$ trials, on average. Consequently, higher code rates lead to a faster decoding when using GRAND because the words' space becomes denser.

\section{System model for coded massive MIMO}
\begin{figure}[t]
   \textit{}
   \centering
    \includegraphics[width=1.0 \columnwidth, clip=true, draft=false]{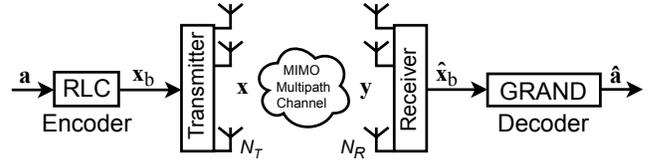}
   \caption{System model for coded mMIMO URLLC.}
    \label{fig:system}
\end{figure}

A coded mMIMO system is considered, with a RLC encoder at the transmitter and GRAND at the receiver, as depicted in Fig.~\ref{fig:system}. The description of the transmission chain is made considering only one packet of information comprising $k$ bits, which is encoded onto a codeword, sent via spatial multiplexing. For longer streams the process can be repeated by slicing the string of bits in packets of size $k$.

\subsection{RLC encoding scheme}

A block $\textbf{a}$ of $k$ i.i.d. information bits is linearly encoded onto a codeword of bits $\textbf{x}_\text{b}$ of length $n$ using a systematic RLC with rate $R=\frac{k}{n}$ over the binary Galois $\mathbb{F}_2$.
A $(n,k)$ RLC defines a codebook $\mathcal{C}$ with $2^k=2^{nR}$ codewords of length $n$, constituting a linear subspace of the discrete vector space $\mathbb{F}_{2}^{n}$. The minimal Hamming distance between two codewords in $\mathcal{C}$ is $d$, however, its role in the context of RLC is not as relevant to determine the performance of some code \cite[Ch.13]{MacKay2003}.
The code is defined by a binary \textit{random} generator matrix $\mathbf{G} \in \mathbb{F}_{2}^{k \times n}$, which acts as the basis matrix  for the code subspace, such that $\mathcal{C}=\left\{\textbf{x}_\text{b}  = \textbf{a} \mathbf{G}: \textbf{a} \in \mathbb{F}_{2}^{k}\right\}$. The generator matrix is of the form $\mathbf{G}=\left[\ \mathbf{P} \ | \ \mathbf{I}_k \ \right]$, where $\mathbf{P} \in \mathbb{F}_{2}^{k \times (n-k)}$ is a binary random matrix, and $\mathbf{I}_k$ is the identity matrix of size $k \times k$, responsible for the systematic part of the encoding. As a result of this construction, all codewords $\mathbf{x}_\text{b}$ are equally probable.

\subsection{Spatial multiplexing with mMIMO}\label{sec:mMIMO}

The baseband bit encoder outputs the $n$-bit codeword $\textbf{x}_\text{b}$, which is fed to a transmitter equipped with $N_T$ transmit antennas which maps the $n$ bits onto $N_T$ complex symbols belonging to a $M$-QAM constellation defined by the symbol alphabet $\mathcal{A \in \mathbb{C}}$, each of which carrying $\log_2(M)$ bits. These symbols form the complex transmit vector is $\mathbf{x}^{(c)} = [x_1, ... , x_{N_T} ]^T $. In this work one uses a ``one-shot'' approach to the transmission of each codeword. This means that the number of transmit antennas fits the number of complex symbols needed to transmit one codeword when using a given  cardinality $M$ for the modulation, and is given by
\begin{equation}
    N_T= \frac{n}{\log_2(M)},\label{antennas}
\end{equation}
nevertheless, this constraint can be easily alleviated.

The codeword of $n$ bits is divided in small strings with $\log_2(M)$ bits, which are mapped  onto $N_T$ constellation symbols using some mapping scheme. Two mapping schemes will be considered: \textit{natural mapping} and \textit{Gray mapping}. In both cases the complex constellation alphabet $\mathcal{A}$ is constructed as the Cartesian product of two pulse amplitude modulation (PAM) alphabets, where each of the I and Q components carry $\frac{\log_2(M)}{2}$ PAM bits. The system is designed such that the fist $n/2$ bits of a codeword control the PAM symbols in the I component, and the remaining $n/2$ bits control the PAM symbols in the Q component. Square QAM constellations with $M=2^{2 \theta}$ are considered, with $\theta=2, 3$.

The symbols transmitted from each antenna are assumed to have unit power, that is $\mathbb{E}\{(\mathbf{x}^{(c)})^H \mathbf{x}^{(c)}\}=M$, where $(.)^H$ is the Hermitian transpose operator. The coded transmitted signal $\mathbf{x}$ goes through a channel (perfectly known at the receiver), characterized by the matrix $\mathbf{H} \in \mathbb{C}^{N_R \times N_T}$ and is received at the receiver equipped with $N_R \gg N_T$ antenna elements.
Using the complex MIMO model for flat Rayleigh fading, the received signal $\mathbf{y}= [y_1, ... , y_{N_R} ]^T$ can be expressed as:
\begin{equation}
	\mathbf{y}^{(c)} = \sqrt{\frac{\text{SNR}}{M}} \mathbf{H}^{(c)}\mathbf{x}^{(c)} + \mathbf{n}^{(c)},
\end{equation}
where $\mathbf{n}^{(c)}= [n_1, ... , n_{N_R} ]^T$ represents the additive noise at the receiver. Both the entries in $\mathbf{H}^{(c)}$ and in $\mathbf{n}$ are i.i.d. random variables taken from a complex normal distribution $\mathcal{CN}(0,1)$.  For implementation purposes, the complex-valued MIMO model was converted to a real-valued one \cite{Monteiro2012}.

While it is well-known that the performance of ZF is very poor in symmetric MIMO ($N_R=N_T$), ZF attains quasi-optimal performance in highly asymmetric MIMO with $N_R \gg N_T$. Matrix inversion may become expensive but very good approximate solutions for the inverse matrix can be obtained via Neumann series \cite{Fast2016}.
Considering $N_R \gg N_T$, ZF incurs no performance loss and boosts the receiver array gain.
ZF detection amounts to applying the Moore-Penrose pseudo-inverse \cite{Monteiro2012}
\begin{equation}\label{eq:Moore-Penrose}
\mathbf{H^+}=\left(\mathbf{H}^H\mathbf{H}\right)^{-1} \mathbf{H}^H, \end{equation}
resulting at the receiver:
\begin{equation}
    \mathbf{H^+}\mathbf{y}=\mathbf{I}_{N_T} \mathbf{x}+\underbrace{\mathbf{H^+}\mathbf{n}}_{\mathbf{u}},\label{eq:ZF}
\end{equation}
where $\mathbf{I}_{N_T}$ is the identity matrix of size $N_T$ and $\mathbf{u} \in \mathbb{C}^{N_T}$ denotes the new noise vector after the ZF filter.

Symbol detection is made via a slicer defined by the shape of the constellation, and finally the bits are demapped from the detected symbols, reconstructing a binary word $\mathbf{y}_\text{b}=\mathbf{x}_\text{b} \oplus \mathbf{e}$, with $\oplus$ denoting the modulo-2 addition. This word is fed to the decoder, which applies GRAND to infer the transmitted word $\mathbf{\hat{x}}_\text{b}$, which was corrupted by the error pattern $\mathbf{e}$ (where $\mathbf{y}_\text{b}$, $\mathbf{x}_\text{b}$, and $\mathbf{e}$, are all strings of $n$-bits).

\subsection{Decoder applying GRAND}

The task of the decoder is to estimate $\mathbf{x}_\text{b}$ given $\mathbf{y}_\text{b}$. The central idea of GRAND is that this task is equivalent to the one of decoding the error pattern $\mathbf{\hat{e}}$ affecting $\mathbf{x}_\text{b}$.

Instead of using the generator matrix $\mathbf{G}$, any linear block code can also be defined by its parity check matrix $\mathbf{H} \in \mathbb{F}_{2}^{(n-k) \times n}$ whose kernel is: $\mathcal{C}=\left\{\mathbf{x}_\text{b} \in \mathbb{F}_{2}^{n}: \mathbf{H x}_\text{b}^{T}=\mathbf{0}\right\}$, reminding that $\mathbf{x}_\text{b} \in \mathcal{C}$. \textit{Note}: the notation of the parity matrix coincides with the one of the MIMO channel matrix; given the very different contexts in which they appear, we have opted to keep the traditional notations.

The \textit{syndrome} of a detected word $\mathbf{y}$ is  $\mathbf{s}(\mathbf{x}_\text{b})=\mathbf{H} \mathbf{x}_\text{b}^{T}=\mathbf{H}\left(\mathbf{x}_\text{b}^{T} \oplus \mathbf{e}^{T}\right)=\mathbf{H} \mathbf{e}^{T}$, which is the $\mathbf{0}$ vector only if $\mathbf{e}$ is $\mathbf{0}$ or $\mathbf{e}$ is a valid codeword; in both cases $\mathbf{y}_\text{b} \in \mathcal{C}$. The syndrome associated with each error pattern is not unique. The number of distinct syndromes is only $2^{(n-k)}$ and the number of possible error patterns is $\sum_t^n \binom{n}{t}$. Therefore, identifying the true error pattern based on $\mathbf{s}(\mathbf{\mathbf{x}_\text{b}})$ leads to highly sub-optimal decoding.

\begin{figure}[t]
    \centering
    \includegraphics[width=0.6 \columnwidth, clip=true, draft=false]{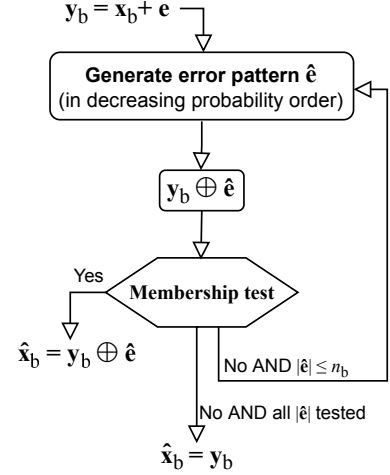}
   \caption{Description of the GRAND algorithm.}
    \label{fig:grand}
\end{figure}

Unlike other deployed decoders, 
GRAND concentrates on identifying the error pattern rather than the codeword itself; it attempts to decode $\mathbf{x}_\text{b}$ by successively testing error patterns in decreasing probability order, as described in Fig.~\ref{fig:grand}.

Because of the combined effect of using a $M$-QAM modulation and having a MIMO flat fading channel, the noise patterns $\mathbf{e}$ will not have the same probability distribution of the additive white Gaussian noise (AWGN) model that has mostly been considered when analysing GRAND. Nevertheless, in this work one considers that the probability of the error patterns gets smaller with an increasing weight, denoted as $|\mathbf{e}|$. Thus, error patterns of weight $|\mathbf{e}|=i$ are tested, with $i=1,\dots, n_\text{b}$, meaning that a maximum number of $n_\text{b}$ bit-flips is considered in a $n$-bit codeword.

\section{Perfect channel hardening lower-bound} \label{sec:CG}
The error patterns $\mathbf{e}$ that corrupt the transmitted codeword are not simply due to the thermal AWGN noise at the receive antennas. As seen in Section \ref{sec:mMIMO}, in  \eqref{eq:ZF}, when the receiver makes ZF detection, the noise affecting the decision is an amplified version of the thermal noise. The effect the ZF filter has on the original noise $\mathbf{n}$ can be tracked by considering the autocorrelation matrix of the new noise $\mathbf{u}= \mathbf{H}^+\mathbf{n}$, which is:
\begin{equation}\label{eq:R_u}
\begin{split}
    \mathbf{R_u} & =\mathbb{E} \left\{ \mathbf{u} \mathbf{u}^H \right\} = \mathbb{E} \left\{ \left(\mathbf{H}^+\mathbf{n} \right) \left(\mathbf{H}^+\mathbf{n} \right)^H \right\}\\
    & = \mathbb{E} \left\{ \left(\mathbf{H}^+\mathbf{n} \right)
    \left(  \mathbf{n}^H (\mathbf{H}^+)^H  \right) \right\}\\
    & =  \mathbf{H}^+  \mathbb{E} \left\{  \mathbf{n} \mathbf{n}^H \right\} (\mathbf{H}^+)^H = \sigma_\text{n}^2 \mathbf{H}^+ (\mathbf{H}^+)^H,
\end{split}
\end{equation}
where the autocorrelation of additive noise, $ \mathbb{E}  \left\{  \mathbf{n} \mathbf{n}^H \right\} = \mathbf{R_n} = \sigma_\text{n}^2 \mathbf{I}_{N_R}$, was used.
Replacing \eqref{eq:Moore-Penrose} in \eqref{eq:R_u}, after some matrix algebra, it is possible to obtain
\begin{equation}
    \mathbf{R_u} = \sigma_\text{n}^2 \left(\mathbf{H}^H\mathbf{H}\right)^{-1}=\sigma_\text{n}^2 \mathbf{T}^{-1},
\end{equation}
where $\mathbf{T}=\textbf{H}^H\textbf{H} \in \mathcal{C}^{N_T \times N_T}$ is the Gram matrix \cite{Monteiro_2010}.

As seen in \eqref{eq:ZF}, the correct detection of $\mathbf{\hat{x}}$ is perturbed by the modified noise vector $\mathbf{u}$.
It is possible to show that the output SNR after ZF detection of the $N_T$ incoming signals is always lower than the input  SNR \cite{Jiang2011} \cite[sec. 2.5.2]{Monteiro2012}:
\begin{equation}
   \mathrm{snr}_i^{(ZF)}=\frac{\mathrm{snr}_i}{\left[\left(\mathbf{H}^{H} \mathbf{H}\right)^{-1}\right]_{ii}}, \quad 1 \leq i \leq N_T,
\end{equation}
 meaning that, in real-world channels, ZF detection always leads to noise enhancement in the detection of $\mathbf{\hat{x}}$. 

There is only one particular (and ideal) situation that does not lead to noise enhancement: when all the column vectors in $\mathbf{H}$ are mutually orthogonal. This is precisely what tends to happen when $N_T$ is fixed and $N_R \rightarrow \infty$, as in the case of mMIMO, leading to the so-called \textit{channel hardening effect}. The geometric interpretation is the following: one has $N_T$ random Gaussian vectors living in a $N_R$-dimensional space; with high probability any pair of the $N_T$ vectors will be orthogonal to each other.
In that case, the Gram matrix, which comprises all the inner products $\mathbf{h}^H_{i}\mathbf{h}_{j}$, $i,j = 1,\dots\ N_T$, becomes a diagonal matrix of the form:
\begin{equation}\label{eq_Gram}
    \mathbf{T}= \text{diag} \big(||h_{11}||^2, \dots, \|h_{N_T N_T}||^2\big)
    = N_R \mathbf{I}_{N_T},
\end{equation}
given that $\|h_{ii}||^2 =\sum_{i=1}^{N_R} |h_i|^2 = N_R$, for all the $N_T$ vectors.

Replacing \eqref{eq_Gram} in \eqref{eq:R_u}, in the case of perfect channel hardening, the autocorrelation of the noise after ZF is $ \mathbf{R_u}=  \frac{\sigma_\text{n}^2}{N_R} \mathbf{I}_{N_T}$.
Finally, the power of $\mathbf{u}$ is $\|\mathbf{u}\|^2=\Tr \left( \mathbf{R_u} \right) = \frac{\sigma_\text{n}^2 N_T}{N_R}$.

It is now possible to establish the equivalent channel model if the $N_T \times N_R$ mMIMO configurations were to attain perfect channel hardening at those (finite) dimensions: $\mathbf{y}=\mathbf{I}_{N_T} \mathbf{x}+\mathbf{u}$.
One has $N_T$ independent parallel channels with each of the  $N_T$ component of $\mathbf{u}$ having power $|u_i|^2=\frac{\sigma_\text{n}^2}{N_R}$, shedding light on the tremendous benefit of having a larger receiver array. 

\section{Results}
The proposed system was assessed by numerical simulation in terms of performance and decoding complexity when transmitting a short codeword in ``one shot'', i.e., only using one MIMO burst, aiming at URLLC. The results are also compared with the (128,103) benchmark polar code used in the control channel of the 5G air interface, thus having $R=0.8$. For an increasing spectral efficiency of the modulation, the number of transmit antennas is reduced to accommodate the same payload of $n$ bits, according to expression \eqref{antennas}. In the case of $M=64$, expression \eqref{antennas} would lead to a non-integer $N_T=21.3$. To make it an integer number of antennas, the RLC code is slightly changed to (132,106) in the case of 64-QAM, keeping the same code rate $R=\frac{106}{132}=0.8$.

The assessment of the system's performance is made via the block error rate (BLER) versus $\frac{E_b}{N_0}$, as it is most common in recent papers assessing GRAND. $E_b$ denotes the energy per information bit and $N_0$ is the bilateral spectral density of the noise at each one of the receive antennas.
The decoding complexity is measured by the expected number of membership tests needed at each $\frac{E_b}{N_0}$. All the results are ergodic in sense that each Monte Carlo iteration uses a different channel matrix $\mathbf{H}$ and a different generator matrix $\mathbf{G}$ for the RLC.

\begin{figure}[t]
    \centering
    \includegraphics[width=1\columnwidth, clip=true, draft=false]{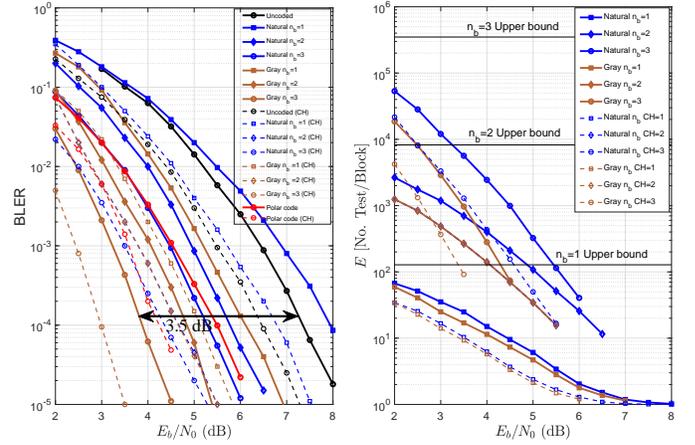}
   \caption{Performance (left) and decoding complexity (right) for different $n_\text{b}$ thresholds in GRAND, using RLC (128,103), with $N_T=32$, and 16-QAM. The corresponding perfect channel hardening lower-bounds are also plotted. The performance of the polar-code (128,103) is also included.}
   \label{fig:16qam}
\end{figure}

\begin{figure}[t]
    \centering
    \includegraphics[width=1\columnwidth, clip=true, draft=false]{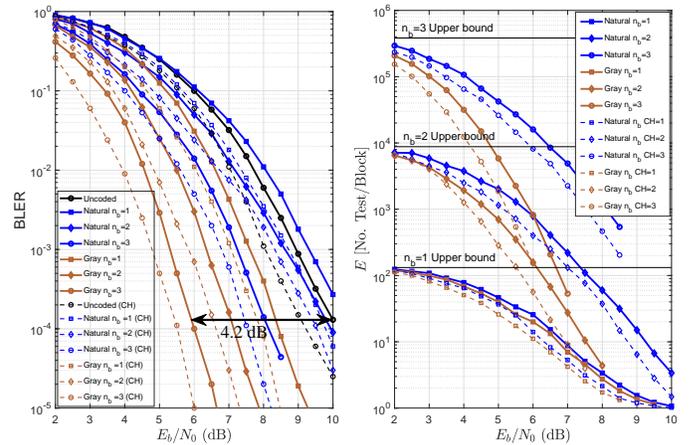}
   \caption{Performance (left) and decoding complexity (right) for different $n_\text{b}$ thresholds in GRAND, using RLC (133,106), with $N_T=22$, and 64-QAM. The corresponding perfect channel hardening lower-bounds are also plotted.}
   \label{fig:64qam}
\end{figure}

Figures~\ref{fig:16qam} and~\ref{fig:64qam} show the performance and complexity results respectively for $M=16, 64$. One should highlight that $n$ remains constant in the different cases, while $N_T$ and $M$ change, with opposite effects and thus the joint effect does not lend itself to an easy interpretation. 
As one would expect, with all other parameters unchanged, Gray mapping always outperforms the equivalent scheme using natural mapping.


\begin{figure}[t]
    \centering
    \includegraphics[width=1\columnwidth, clip=true, draft=false]{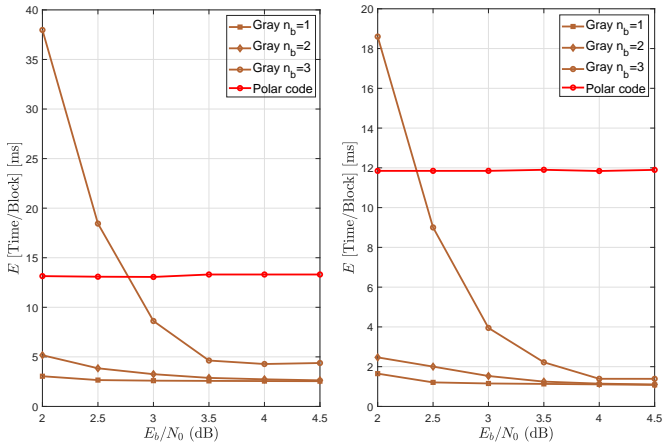}
   \caption{Comparison of the average decoding time per codeword: RLC (128,103) using GRAND with different $n_\text{b}$ thresholds vs the equivalent (128,103) polar code using list decoding. The results are for $N_T=64$ with 4-QAM (left), and $N_T=32$ with 16-QAM (right).}
   \label{fig:polar}
\end{figure}

By increasing the value of $n_\text{b}$, threshold error patterns of larger weight are tested and therefore the performance dramatically increases, at the expense of a much larger member of membership tests. The upper bound for the number of membership tests is given by
\begin{equation}\label{eq:upper_bound}
    UB= \sum_{t=1}^{n_\text{b}} \binom{n}{t} = \binom{n}{1} + \binom{n}{2} + \dots + \binom{n}{n_\text{b}} ,
\end{equation}
however, the results show that the average number of membership tests is much lower than \eqref{eq:upper_bound} in the case $M=16$ (and also with $M=4$, not shown due to space limitations). With 64-QAM the complexity can approach the upper bound when the noise is too large due to the sheer number of modulation symbols in error.
It is interesting to observe how the average number of membership tests always tends to one, which is a consequence of the decreasing number of bits in error, such that eventually almost all received words are valid codewords.

The comparison of the performance of the proposed schemes using RLC and GRAND with the performance attained using the polar code (128,103) decoded using the list decoding technique used in 5G \cite{TalVardy2015}, for a list length of 8, shows that the RLC with GRAND remarkably outperforms the polar code. More surprisingly, this is true even when GRAND runs with only $n_\text{b}=2$ (seen in Fig.~\ref{fig:16qam} and also with 4-QAM, not shown due to space limitations).
Note that for $M=64$ one could easily construct a RLC with a rate that fits the setup, however, a polar code counterpart does not exist and could not be included for comparison in Fig.~\ref{fig:64qam}).

Figures \ref{fig:16qam} and \ref{fig:64qam} include both the performance and the complexity results for the situation with perfect channel hardening (see Sec. \ref{sec:CG}). This gauges the loss coming form the non-ideal mMIMO. As expected, the more asymmetrical the MIMO configuration becomes, the closer one is from that ideal case. One has a fixed $N_R=200$ and $N_T= 32, 22$. In fact, in the $(M=64, N_T=22)$ setup the gap to the ideal channel is only $\approx 0.5$ dB for all $n_\text{b}$ and for both mapping schemes.

The complexity involved in decoding the polar code is assessed by means of the average decoding time of codewords, as depicted in Fig.~\ref{fig:polar}. The decoding time of polar codes remains constant for all $\frac{E_b}{N_0}$. However, as a consequence of the accentuated reduction of the number of membership tests performed (as observed in figures \ref{fig:16qam}, and \ref{fig:64qam}), GRAND almost always spends much less time decoding a codeword than the list decoder. Depending on $\frac{E_b}{N_0}$, when GRAND is set with $n_\text{b}=1,2$ it is $\approx 2.5$ to $\approx 5$ times faster in the $(M=4, N_T=64)$ setup and $\approx 5$ to $\approx 12$ times faster in the $(M=16, N_T=32)$ setup.
Even with the best performant $n_\text{b}=3$, applying GRAND to the RLC is faster than decoding the polar code for $\frac{E_b}{N_0}>2.75$ dB in the first setup and $>2.3$ dB in the second. Using RLC with GRAND is up to over $\approx 3$ times faster in the first setup and up to $\approx 8$ times faster in the second setup. The average decoding time was measured using MATLAB on PC equipped with a CPU Intel Core i7-12700 and 32GB of RAM with maximum clock speed of 4.9 GHz.

\section{Conclusions}
This paper proposed transmission schemes for high-throughput, high-reliability, and very-low latency, adequate for URLLC. The proposal puts together RLCs using GRAND transmitted over mMIMO. The coded schemes attain a large gain ($3.5-4.7$ dB) in respect to the uncoded transmission, significantly outperform the polar-coded counterparts with state-of-the-art list decoding, and for most of the SNR range GRAND delivers a much faster decoding time, thus reducing decoding latency, in addition to the delay savings for not requiring the use of interleavers. Performance can be traded-off with decoding time, and one can get decoding speeds up to 12 times faster than the schemes with polar codes, while still beating the performance of the polar code.

\section*{Acknowledgments}
This work has been funded by Instituto de Telecomunica\c{c}\~{o}es and FCT/MCTES (Portugal) through national funds and when applicable co-funded EU funds under the project UIDB/50008/2020. Sahar Allahkaram is funded by ISCTE-IUL with a Merit Scholarship awarded by the ISCTE School of Technology and Architecture (ISTA).

\bibliographystyle{IEEEtran}
\bibliography{library.bib}

\begin{thebibliography}{10}
\providecommand{\url}[1]{#1}
\csname url@samestyle\endcsname
\providecommand{\newblock}{\relax}
\providecommand{\bibinfo}[2]{#2}
\providecommand{\BIBentrySTDinterwordspacing}{\spaceskip=0pt\relax}
\providecommand{\BIBentryALTinterwordstretchfactor}{4}
\providecommand{\BIBentryALTinterwordspacing}{\spaceskip=\fontdimen2\font plus
\BIBentryALTinterwordstretchfactor\fontdimen3\font minus
  \fontdimen4\font\relax}
\providecommand{\BIBforeignlanguage}[2]{{%
\expandafter\ifx\csname l@#1\endcsname\relax
\typeout{** WARNING: IEEEtran.bst: No hyphenation pattern has been}%
\typeout{** loaded for the language `#1'. Using the pattern for}%
\typeout{** the default language instead.}%
\else
\language=\csname l@#1\endcsname
\fi
#2}}
\providecommand{\BIBdecl}{\relax}
\BIBdecl

\bibitem{Nouri2020}
P.~Nouri, H.~Alves, M.~A. Uusitalo, O.~{Alcaraz L{\'{o}}pez}, and M.~Latva-aho,
  ``{Machine-type wireless communications enablers for beyond 5G: Enabling
  URLLC via diversity under hard deadlines},'' \emph{Computer Networks}, vol.
  174, no.~3, p. 107227, 2020.

\bibitem{An2022}
W.~An, M.~Médard, and K.~R. Duffy, ``Keep the bursts and ditch the
  interleavers,'' \emph{IEEE Transactions on Communications}, vol.~70, no.~6,
  pp. 3655--3667, 2022.

\bibitem{Abbas}
\BIBentryALTinterwordspacing
S.~M. Abbas, M.~Jalaleddine, and W.~J. Gross, ``{GRAND} for {R}ayleigh fading
  channels.'' [Online]. Available: \url{https://arxiv.org/abs/2205.00030}
\BIBentrySTDinterwordspacing

\bibitem{Shannon1948}
C.~E. Shannon, ``A mathematical theory of communication,'' \emph{The Bell
  System Technical Journal}, vol.~27, pp. 379--423, 1948.

\bibitem{Gallager1973}
R.~Gallager, ``The random coding bound is tight for the average code
  (corresp.),'' \emph{IEEE Transactions on Information Theory}, vol.~19, no.~2,
  p. 244–246, Mar 1973.

\bibitem{Polyanskiy_2010}
Y.~Polyanskiy, H.~V. Poor, and S.~Verdu, ``Channel coding rate in the finite
  blocklength regime,'' \emph{IEEE Transactions on Information Theory},
  vol.~56, no.~5, pp. 2307--2359, 2010.

\bibitem{Polyanskiy_2014}
W.~Yang, G.~Durisi, T.~Koch, and Y.~Polyanskiy, ``Quasi-static multiple-antenna
  fading channels at finite blocklength,'' \emph{IEEE Transactions on
  Information Theory}, vol.~60, no.~7, pp. 4232--4265, 2014.

\bibitem{Duffy2019}
K.~R. Duffy, J.~Li, and M.~Médard, ``Capacity-achieving guessing random
  additive noise decoding,'' \emph{IEEE Transactions on Information Theory},
  vol.~65, no.~7, pp. 4023--4040, Jul. 2019.

\bibitem{Duffy2022}
K.~R. Duffy, M.~Médard, and W.~An, ``Guessing random additive noise decoding
  with symbol reliability information ({SRGRAND}),'' \emph{IEEE Transactions on
  Communications}, vol.~70, no.~1, pp. 3--18, 2022.

\bibitem{Duffy2020}
K.~R. Duffy, A.~Solomon, K.~M. Konwar, and M.~Médard, ``{5G NR CA}-polar
  maximum likelihood decoding by {GRAND},'' \emph{54th Annual Conf. on Info.
  Sciences and Systems (CISS)}, 2020.

\bibitem{Duffy2021}
K.~R. Duffy, ``Ordered reliability bits guessing random additive noise
  decoding,'' in \emph{Proc. of IEEE Inter. Conf. on Acoustics, Speech and
  Signal Processing (ICASSP)}, Toronto, Canada, Jun 2021, p. 8268–8272.

\bibitem{MacKay2003}
D.~J.~C. MacKay, \emph{Information Theory, Inference and Learning
  Algorithms}.\hskip 1em plus 0.5em minus 0.4em\relax Cambridge, UK: Cambridge
  University Press, 2003.

\bibitem{Monteiro2012}
F.~Monteiro, ``Lattices in {MIMO} spatial multiplexing: Detection and
  geometry,'' Ph.D. dissertation, University of Cambridge, United Kingdom,
  2012.

\bibitem{Fast2016}
F.~Rosário, F.~A. Monteiro, and A.~Rodrigues, ``Fast matrix inversion updates
  for massive {MIMO} detection and precoding,'' \emph{IEEE Signal Processing
  Letters}, vol.~23, no.~1, pp. 75--79, 2016.

\bibitem{Monteiro_2010}
F.~A. Monteiro and I.~J. Wassell, ``Recovery of a lattice generator matrix from
  its {Gram} matrix for feedback and precoding in {MIMO},'' in \emph{2010 4th
  Inter. Symp. on Comm., Control and Sig. Proc. (ISCCSP)}, 2010.

\bibitem{Jiang2011}
Y.~Jiang, M.~K. Varanasi, and J.~Li, ``Performance analysis of {ZF} and {MMSE}
  equalizers for {MIMO} systems: An in-depth study of the high {SNR} regime,''
  \emph{IEEE Transactions on Information Theory}, vol.~57, no.~4, pp.
  2008--2026, 2011.

\bibitem{TalVardy2015}
I.~Tal and A.~Vardy, ``List decoding of polar codes,'' \emph{IEEE Transactions
  on Information Theory}, vol.~61, no.~5, pp. 2213--2226, 2015.

\end{thebibliography}
\end{document}